\documentclass[aps,prc,twocolumn,superscriptaddress,showpacs,floatfix]{revtex4-1}
\usepackage{color}
\usepackage{subfigure}
\usepackage{amsmath,epsfig}
\usepackage{graphicx}
\usepackage{epstopdf}
\usepackage{dcolumn}
\usepackage{bm}
\usepackage{multirow}
\usepackage{hhline}
\usepackage{afterpage}
\usepackage{hyperref} 
\usepackage{float}
\usepackage{booktabs}

\begin{document}


\title{Azimuthal anisotropy measurement of (multi-)strange hadrons in Au+Au collisions at $\sqrt{s_{\text{NN}}}$ = 54.4 GeV}
\affiliation{American University of Cairo, New Cairo 11835, New Cairo, Egypt}
\affiliation{Texas A\&M University, College Station, Texas 77843}
\affiliation{Brookhaven National Laboratory, Upton, New York 11973}
\affiliation{AGH University of Science and Technology, FPACS, Cracow 30-059, Poland}
\affiliation{Ohio State University, Columbus, Ohio 43210}
\affiliation{University of Kentucky, Lexington, Kentucky 40506-0055}
\affiliation{Panjab University, Chandigarh 160014, India}
\affiliation{Variable Energy Cyclotron Centre, Kolkata 700064, India}
\affiliation{Abilene Christian University, Abilene, Texas   79699}
\affiliation{Instituto de Alta Investigaci\'on, Universidad de Tarapac\'a, Arica 1000000, Chile}
\affiliation{University of California, Riverside, California 92521}
\affiliation{University of Houston, Houston, Texas 77204}
\affiliation{University of Jammu, Jammu 180001, India}
\affiliation{State University of New York, Stony Brook, New York 11794}
\affiliation{Czech Technical University in Prague, FNSPE, Prague 115 19, Czech Republic}
\affiliation{Nuclear Physics Institute of the CAS, Rez 250 68, Czech Republic}
\affiliation{Shanghai Institute of Applied Physics, Chinese Academy of Sciences, Shanghai 201800}
\affiliation{Yale University, New Haven, Connecticut 06520}
\affiliation{University of California, Davis, California 95616}
\affiliation{Lawrence Berkeley National Laboratory, Berkeley, California 94720}
\affiliation{University of California, Los Angeles, California 90095}
\affiliation{Indiana University, Bloomington, Indiana 47408}
\affiliation{Warsaw University of Technology, Warsaw 00-661, Poland}
\affiliation{Shandong University, Qingdao, Shandong 266237}
\affiliation{Fudan University, Shanghai, 200433 }
\affiliation{University of Science and Technology of China, Hefei, Anhui 230026}
\affiliation{Tsinghua University, Beijing 100084}
\affiliation{University of California, Berkeley, California 94720}
\affiliation{ELTE E\"otv\"os Lor\'and University, Budapest, Hungary H-1117}
\affiliation{University of Heidelberg, Heidelberg 69120, Germany }
\affiliation{Wayne State University, Detroit, Michigan 48201}
\affiliation{Indian Institute of Science Education and Research (IISER), Berhampur 760010 , India}
\affiliation{Kent State University, Kent, Ohio 44242}
\affiliation{Rice University, Houston, Texas 77251}
\affiliation{University of Tsukuba, Tsukuba, Ibaraki 305-8571, Japan}
\affiliation{University of Illinois at Chicago, Chicago, Illinois 60607}
\affiliation{Lehigh University, Bethlehem, Pennsylvania 18015}
\affiliation{University of Calabria \& INFN-Cosenza, Italy}
\affiliation{National Cheng Kung University, Tainan 70101 }
\affiliation{Purdue University, West Lafayette, Indiana 47907}
\affiliation{Southern Connecticut State University, New Haven, Connecticut 06515}
\affiliation{Central China Normal University, Wuhan, Hubei 430079 }
\affiliation{Technische Universit\"at Darmstadt, Darmstadt 64289, Germany}
\affiliation{Temple University, Philadelphia, Pennsylvania 19122}
\affiliation{Valparaiso University, Valparaiso, Indiana 46383}
\affiliation{Indian Institute of Science Education and Research (IISER) Tirupati, Tirupati 517507, India}
\affiliation{Institute of Modern Physics, Chinese Academy of Sciences, Lanzhou, Gansu 730000 }
\affiliation{Frankfurt Institute for Advanced Studies FIAS, Frankfurt 60438, Germany}
\affiliation{National Institute of Science Education and Research, HBNI, Jatni 752050, India}
\affiliation{University of Texas, Austin, Texas 78712}
\affiliation{Rutgers University, Piscataway, New Jersey 08854}
\affiliation{Institute of Nuclear Physics PAN, Cracow 31-342, Poland}
\affiliation{Max-Planck-Institut f\"ur Physik, Munich 80805, Germany}
\affiliation{Creighton University, Omaha, Nebraska 68178}
\affiliation{Indian Institute Technology, Patna, Bihar 801106, India}
\affiliation{Ball State University, Muncie, Indiana, 47306}
\affiliation{Universidade de S\~ao Paulo, S\~ao Paulo, Brazil 05314-970}
\affiliation{Huzhou University, Huzhou, Zhejiang  313000}
\affiliation{Michigan State University, East Lansing, Michigan 48824}
\affiliation{Argonne National Laboratory, Argonne, Illinois 60439}
\affiliation{United States Naval Academy, Annapolis, Maryland 21402}
\affiliation{South China Normal University, Guangzhou, Guangdong 510631}

\author{M.~S.~Abdallah}\affiliation{American University of Cairo, New Cairo 11835, New Cairo, Egypt}
\author{B.~E.~Aboona}\affiliation{Texas A\&M University, College Station, Texas 77843}
\author{J.~Adam}\affiliation{Brookhaven National Laboratory, Upton, New York 11973}
\author{L.~Adamczyk}\affiliation{AGH University of Science and Technology, FPACS, Cracow 30-059, Poland}
\author{J.~R.~Adams}\affiliation{Ohio State University, Columbus, Ohio 43210}
\author{J.~K.~Adkins}\affiliation{University of Kentucky, Lexington, Kentucky 40506-0055}
\author{I.~Aggarwal}\affiliation{Panjab University, Chandigarh 160014, India}
\author{M.~M.~Aggarwal}\affiliation{Panjab University, Chandigarh 160014, India}
\author{Z.~Ahammed}\affiliation{Variable Energy Cyclotron Centre, Kolkata 700064, India}
\author{D.~M.~Anderson}\affiliation{Texas A\&M University, College Station, Texas 77843}
% ...

\collaboration{STAR Collaboration}\noaffiliation

\begin{abstract}
Azimuthal anisotropy of produced particles is one of the most important observables used to access the collective properties of the expanding medium created in relativistic heavy-ion collisions. In this paper, we present second ($v_{2}$) and third ($v_{3}$) order azimuthal anisotropies of $K_{S}^{0}$, $\phi$, $\Lambda$, $\Xi$ and $\Omega$ at mid-rapidity ($|y|<$1) in Au+Au collisions at $\sqrt{s_{\text{NN}}}$ = 54.4 GeV measured by the STAR detector. 
The $v_{2}$ and $v_{3}$  are measured as a function of transverse momentum and centrality. Their energy dependence is also studied. $v_{3}$ is found to be more sensitive to the change in the center-of-mass energy than $v_{2}$. Scaling by constituent quark number is found to hold  for $v_{2}$ within 10\%. This observation could be evidence for the development of partonic collectivity in 54.4 GeV Au+Au collisions. Differences in $v_{2}$ and $v_{3}$ between baryons and anti-baryons are presented, and ratios of $v_{3}$/$v_{2}^{3/2}$ are studied and motivated by hydrodynamical calculations. The ratio of $v_{2}$ of $\phi$ mesons to that of anti-protons ($v_{2}(\phi)/v_{2}(\bar{p})$) shows centrality dependence at low transverse momentum, presumably resulting from the larger effects from hadronic interactions on anti-proton $v_{2}$.
\end{abstract}

\maketitle

\section{Introduction}
According to Quantum ChromoDynamics (QCD), at very high temperature ($\rm{T}$) and/or large baryonic chemical potential ($\mu_{B}$) a deconfined phase of quarks and gluons is expected to be present, while at low $\rm{T}$ and low $\mu_{B}$ quarks and gluons are known to be confined inside hadrons~\cite{QCD}.  High energy heavy-ion collisions provide a unique opportunity to study QCD matter at extremely high temperature and density.  
 Experiments at the Relativistic Heavy Ion Collider (RHIC) have shown that a very dense medium  of deconfined quarks and gluons is formed in Au+Au collisions at the center-of-mass energy of $\sqrt{s_{\text{NN}}}$ = 200 GeV~\cite{early_freezeout,star_200_1,star_200_2,star_200_3,star_200_4,phenix_200_1,phenix_200_2,phobos_200}.  Azimuthal anisotropy parameters ($v_{n}$), which quantify the azimuthal asymmetries of particle production in momentum space, are an excellent tool to study the properties of the deconfined medium created in these collisions~\cite{flow1,flow2,flow3,flow4,flow5,flow6,flow7,flow8}. Observations of large $v_{n}$ magnitudes and their constituent quark scaling in 200 GeV Au+Au collisions ($\mu_{B}$ $\sim$ 20 MeV) have been considered a signature of partonic collectivity of the system~\cite{star_200gev_prl}.
 \par
To study the QCD phase structure over a large range in $\rm{T}$ and $\mu_{B}$, a beam energy scan program has been carried out by RHIC~\cite{rhic_bes}.  The first phase of this program (BES-I) was carried out in 2010-14.
Measurements of azimuthal anisotropies of light flavor hadrons made during during the BES-I program by the STAR experiment indicate the formation of  QCD matter dominated by hadronic interactions in Au+Au collisions at $\sqrt{s_{\text{NN}}}$ $<$ 11.5 GeV ($\mu_{B}$ $>$ 200 MeV)~\cite{star_bes_prl,bes_pid_prc}.
\par
Strange hadrons, especially those containing more than one strange quark, are considered a good probe to study the collective properties of the medium created in the early stage of heavy-ion collisions~\cite{small_x_1,sss,early_freezeout,BM_SQM2008,J_Chen,MD_BM}. The measurement of average transverse momentum $\left<p_{T}\right>$ of  $\phi$ mesons shows weak centrality dependence while $\left<p_{T}\right>$  of protons increases significantly from peripheral to central collisions. This could be due to the fact that $\phi$ mesons have relatively small hadronic interaction cross-section compared to that of proton~\cite{phi_502TeV}. 
 Measurements of (multi-)strange hadron $v_{n}$ is limited by the available statistics in BES-I.
In this paper, we report high precision measurements of azimuthal anisotropy parameters, $v_{2}$ and $v_{3}$, of strange and multi-strange hadrons at mid-rapidity ($|y| <$ 1) in Au+Au collisions at $\sqrt{s_{\text{NN}}}$ = 54.4 GeV ($\mu_{B}$ $\sim$ 90 MeV). 
$v_{2}$ and $v_{3}$ of K$_{S}^{0}$, $\phi$, $\Lambda$, $\Xi$ and $\Omega$ are measured as a function of particle transverse momentum ($p_{T}$) and collision centrality. Such measurements will provide deep insights into properties of the hot and dense medium, such as partonic collectivity, transport coefficients, and hadronization mechanisms. 
\par
This paper is organized in the following manner.  In sections II, III and IV, we describe the dataset, the analysis method, and systematic studies respectively. In section V  we report the results. Finally, a summary is given in section VI.

\section{Experimental setup}
\label{STAR detectors and experimental setup}
In this analysis, a total of 600 M minimum bias Au+Au events at $\sqrt{s_{\text{NN}}}$ = 54.4 GeV recorded by the STAR experiment are used. Events for analysis are selected based on the collision vertex position. Along the beam direction, a vertex position cut of $|V_{z}|$ $<$ 30 cm is applied. A radial vertex position cut (defined as $V_{r}$ =$\sqrt{V_{x}^{2} +V_{y}^{2}}$) of $V_{r}$ $<$ 2.0 cm is used in order to avoid collision with beam pipe whose radius is 3.95 cm.  

The trajectory of a charged particle through STAR’s magnetic field can be reconstructed, and thus its momentum determined, using the Time Projection Chamber (TPC)~\cite{tpc}.
To ensure good track quality, the number of TPC hit points on each track is required to be larger than 15, and the ratio of the number of used TPC hit points to the maximum possible number of hit points along the trajectory should be larger than 0.52. 
The transverse momentum of each particle is limited to $p_{T}$ $>$ 0.15 GeV/c.
\begin{figure}[!htbp]
\centering
\includegraphics[scale=0.4]{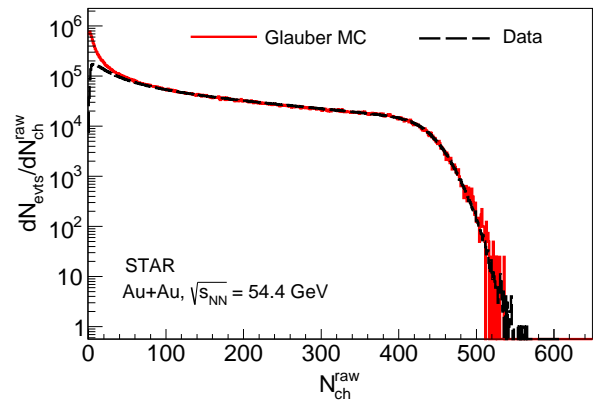}
\caption{ The uncorrected multiplicity distribution of reconstructed charged particles in Au+Au collisions at $\sqrt{s_{\text{NN}}}$ = 54.4 GeV. Glauber Monte Carlo simulation is shown as the solid red curve. }
\label{fig:refmult}
\end{figure}
\par
The collision centrality is determined by comparing the uncorrected charged particle multiplicity within a pseudorapidity range of $|\eta|$ $<$ 0.5 measured by the TPC with a Glauber Monte Carlo (MC)~\cite{glauber} simulation as shown in Fig.~\ref{fig:refmult}. The significant difference between the measured multiplicity and Glauber simulation at low multiplicity values is due to trigger and primary vertex finding inefficiency. This is corrected by taking the ratio of the simulated multiplicity distribution to that in data as a weight factor. The detailed procedure to obtain the simulated multiplicity distribution using Glauber MC is similar to that described in Ref.~\cite{star_bes_chg}. Central (peripheral) events correspond to collisions of large (small) nuclear overlap and thus large (small) charged particle multiplicities.
\par
Particle identification is done using the TPC and the Time-of-Flight (TOF) detectors~\cite{tof} at mid-pseudorapidity ($|\eta|$ $<$1.0).  Both the TPC and TOF have full azimuth coverage.
Long-lived charged particles, e.g. $\pi$, $K$, and $p$, are identified directly using specific ionization energy loss in the TPC and time of flight information in TOF~\cite{bes_pid_prc}. Short-lived strange hadrons ($K_{S}^{0}$, $\phi$, $\Lambda$, $\Xi$, $\Omega$) are reconstructed through two-body hadronic decay channels: $K_{S}^{0}$ $\longrightarrow$ $\pi^{+}$ + $\pi^{-}$, $\phi$ $\longrightarrow$ $\it{K}^{+}$ + $\it{K}^{-}$, $\Lambda (\bar{\Lambda})$ $\longrightarrow$ $\it{p} (\bar{\it{p}})$ + $\pi^{-}(\pi^{+})$, $\Xi^{\pm}$ $\longrightarrow$ $\Lambda$ +$\pi^{\pm}$ and $\Omega^{\pm}$ $\longrightarrow$ $\Lambda$ + $\it{K}^{\pm}$. $K_{S}^{0}$, $\Lambda$, $\Xi$, and $\Omega$ decay weakly and therefore decay topology cuts are applied to reduce the combinatorial background. Cuts on the following topological variables are used:
(1) Distance of Closest Approach (DCA) between the two daughter tracks, (2) the DCA of the daughter tracks to the collision vertex, (3) the DCA of the reconstructed parent strange hadron to the collision vertex, (4) the decay length of the strange hadrons, and (5) the angle between the spatial vector pointing from the collision vertex to the decay vertex and the momentum vector of the parent strange hadron. Since the $\phi$ meson decays strongly, its daughter kaons appear to originate from the collision vertex.  The DCAs of kaon tracks from the collision vertex are required to be less than 3 cm for $\phi$ meson reconstruction.
\par
\begin{figure*}
\centering
\includegraphics[scale=0.7]{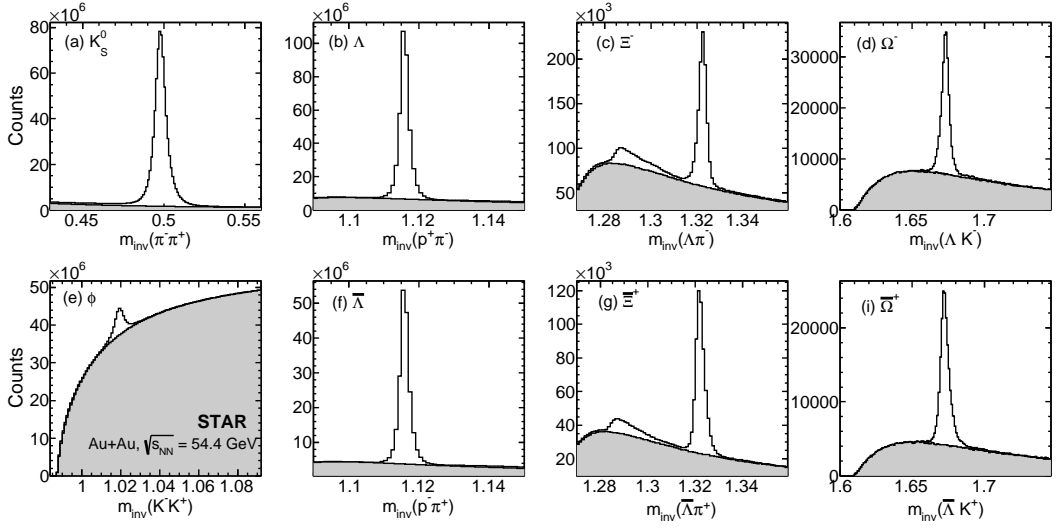}
\caption{Invariant mass distributions  for  $K_{S}^{0}$, $\phi$, $\Lambda$, $\Xi^{-}$, $\Omega^{-}$ and their anti-particles in  minimum bias Au+Au collisions at $\sqrt{s_{\text{NN}}}$ = 54.4 GeV. The combinatorial background is shown as gray shaded histograms. No background subtraction was included in any of the 8 panels.}
\label{fig:massplot}
\end{figure*}

An event mixing technique is used for the subtraction of combinatorial background for the $\phi$ mesons~\cite{mix_evnt1} and different polynomial functions (1$^{st}$ and 2$^{nd}$ order) are used to fit  the background after mixed-event background subtraction. For $K_{S}^{0}$ and $\Lambda$, the like-sign method is used to estimate the background and for $\Xi$ and $\Omega$, the rotational background method is used~\cite{multi_reco_1,multi_reco_2,multi_reco_4}. The invariant mass distributions of  $K_{S}^{0}$, $\phi$, $\Lambda$, $\Xi^{-}$, $\Omega^{-}$ and their anti-particles are shown in Fig.~\ref{fig:massplot}. The invariant mass distribution for $\Xi^{-}$ ($\bar{\Xi}^{+}$) has a small bump due to the combinatorial $\Lambda$ background~\cite{multi_reco_1}.

\section{Analysis method}
\label{Analysis method}
The $n^{th}$ order flow coefficient with respect to the event plane is given by
\begin{equation}
v_{n} = \frac{\langle \cos n(\Phi_{i} - \psi_{n})\rangle}{R_{n}},
\end{equation}
where the angle-bracket represents the average over all the particles in each event and over all the events, $\Phi_{i}$ is the azimuthal angle of the $i^{th}$ particle in an event and $\psi_{n}$ is the event plane angle for the $n^{th}$ order anisotropy of an event~\cite{a_poskazer}. The $R_{n}$ denotes the resolution of the $n^{th}$ order event plan angle. The event plane angle can be determined based on the azimuthal distribution of particles in the plane transverse to the collision axis. The $n^{th}$ order event plane angle is given by
\begin{equation}
\psi_{n} = \frac{1}{n}\tan^{-1} \frac{\sum_{i}w_{i}\sin(n\Phi_{i})}{\sum_{i}w_{i}\cos(n\Phi_{i})}.
\end{equation}
Here $w_{i}$ is the weight factor taken as $p_{T}$ of the particle for optimal resolution. The $n^{th}$ order event plane has a symmetry of 2$\pi$/n and one would expect an isotropic distribution of the event plane angle from 0 to 2$\pi$/n. However, due to the azimuthally non-uniform detection efficiency of the TPC, the reconstructed event plane angle distribution is usually not isotropic. This is corrected for using the $\Phi$-weight method, details of which can be found in the ref.~\cite{a_poskazer}.
\par

To suppress the auto-correlation between particles of interest and those used for event plane angle determination ~\cite{star_bes_chg,a_poskazer}, calculations of the $v_{n}$ coefficients for particles in the positive pseudorapdity region (0 $< $ $\eta$ $<$ 1) utilize the sub-event plane determined using particles in the negative pseudorapdity region (-1 $<$ $ \eta$ $<$ -0.05), and vice versa. Its definition is the following: 
\begin{equation}
\label{eqn:5}
v_{n} = \frac{\langle\cos n (\Phi_{i} - \psi^{A/B}_{n})\rangle}{R_{n}},
\end{equation}
where $\psi^{A}_{n}$ and $\psi^{B}_{n}$ are the sub-event planes in negative (-1 $<$ $ \eta$ $<$ -0.05) and positive (0.05 $ < $ $\eta$ $ < $1) pseudorapidity regions, respectively. In addition to that, auto-correlation has been removed in the case when decay daughters are distributed in sub-events.\\
The event plane resolution $R_{n}$ is estimated using: 
\begin{equation}
R_{n} = \langle \cos n(\psi_{n}-\psi_{R})\rangle = \sqrt{\langle\cos n (\psi^{A}_{n} - \psi^{B}_{n})\rangle},
\end{equation}
in which $\psi_{R}$ is the reaction plane angle. Resolution corrections for wide centrality bins are done using the method described in Ref.~\cite{res_wide}. $\psi_{2}$ and $\psi_{3}$ resolution in different centrality bins are given in Table~\ref{tab_res}

\begin{table}[]
\begin{tabular}{|l|l|l|l|l|}
\hline
Centrality & $\psi_{2}$ resolution  & $\psi_{3}$ resolution \\ \hline
0-5\% & 0.3462 $\pm$ 0.0002 &  0.2284 $\pm$ 0.0003  \\ \hline
5-10\% & 0.4549 $\pm$ 0.0001   & 0.2360 $\pm$ 0.0002    \\ \hline
10-20\% & 0.54179 $\pm$ 0.00007 & 0.2257 $\pm$ 0.0002  \\ \hline
20-30\% & 0.56211 $\pm$ 0.00007  &   0.1981 $\pm$ 0.0002  \\ \hline
30-40\% & 0.51865 $\pm$ 0.00008  &  0.1636 $\pm$ 0.0003 \\ \hline
40-50\% & 0.4338 $\pm$ 0.0001 &  0.1234 $\pm$ 0.0003 \\ \hline
50-60\% & 0.3289 $\pm$ 0.0001 &  0.0863 $\pm$ 0.0005 \\ \hline
60-70\% & 0.2295 $\pm$ 0.0002 &  0.0564 $\pm$ 0.0008 \\ \hline
70-80\% & 0.1578  $\pm$ 0.0003&  0.028 $\pm$ 0.002  \\ \hline
\end{tabular}
\caption{Resolution for $\psi_{2}$ and $\psi_{3}$ in different centrality bins.}
\label{tab_res}
\end{table}


\par
\begin{figure}[!htbp]
\centering
\includegraphics[width=0.4\textwidth]{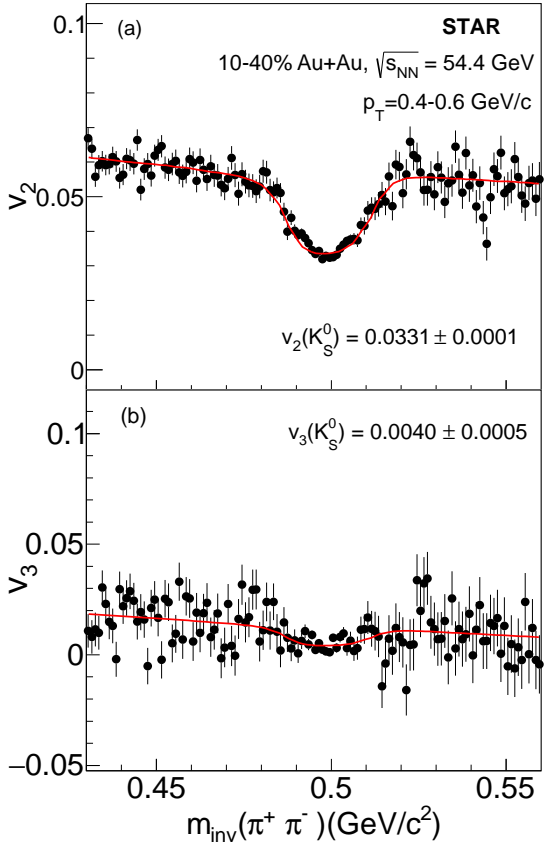}
\caption{The upper panel shows $v_{2}$ as a function of the invariant mass of $\pi^{+}$ $\pi^{-}$ pairs and the lower panel shows the same for $v_{3}$. Red lines represent fit functions given in Eq.~\ref{eqn:6}}
\label{fig:b}
\end{figure}

By using equation ~\ref{eqn:5}, one can calculate the $v_{n}$ of particles that are detected directly and whose azimuthal distributions are known in every event. But the particles used in this analysis are short-lived and can't be detected directly. To calculate the $v_{n}$ of such particles, the invariant mass method is used ~\cite{inv_mass_v2}, in which the $v_{n}$ of the particle of interest is calculated as a function of the invariant mass of the decayed daughter particles. Figure~\ref{fig:b}, taking $K^{0}_{S}$ as an example, shows $v_{2}$ and $v_{3}$ as a function of the $\pi^{+}\pi^{-}$ pair invariant mass in the 10-40\% centrality bin.
The total $v_{n}$ of the signal+background can be decomposed into two parts.
\begin{equation}
\label{eqn:6}
v_{n}^{S+B} = v_{n}^{S} \frac{S}{S+B} + v_{n}^{B} \frac{B}{S+B}.
\end{equation}
Here $v_{n}^{S}$ is the $v_{n}$ of the signal ($K_{S}^{0}$), $v_{n}^{B}$ is the $v_{n}$ of the background,  S is the raw signal counts and B is the background counts. $v_{n}^{B}$ is approximated with a first order polynomial function. $v_{n}^{S}$ is a free parameter and can be obtained by fitting $v_{n}$ using Eq.~\ref{eqn:6}, shown as solid red lines in Fig.~\ref{fig:b}. The $v_{2}$ and $v_{3}$ of other strange hadrons are calculated in a similar way except for $\Xi$. For $\Xi$, Eq.~\ref{eqn:6} has been modified as follows:
\begin{equation}
\label{eqn:7}
v_{n}^{S+B} = v_{n}^{S} \frac{S}{S+B+b} + v_{n}^{b} \frac{b}{S+B+b} + v_{n}^{B} \frac{B}{S+B+b},
\end{equation}
where $b$ denotes the yield of the residual bump observed in the low invariant mass region (see Fig.~\ref{fig:massplot}), and  $v_{n}^{b}$ denotes the $v_{n}$  of the residual candidates in the bump region.
Systematic checks have been carried out to examine the effect of the bump in $\Xi$ $v_{n}$ extraction by changing fit ranges and the shape of the background $v_{n}^{b}$ at the bump region. The effect is found to be negligibly small, less than 1\%, on the $v_{n}$ values of $\Xi$ particles.  
 
\section{Systematic uncertainty}
Systematic uncertainties are evaluated by varying event selection cuts, track selection cuts, and background subtraction methods. Track selection cuts used for event plane angle calculation are also varied.
For particles like $\Xi$ and $\Omega$ the default background construction method is the rotational method and for particles like $K_{S}^{0}$ and $\Lambda$ the default background construction method is the like-sign method. 
As an alternative to estimate the background fraction, polynomial functions are used to model the residual background in fitting the invariant mass distributions. 
The resulting differences in $v_{n}$ between using the default and alternative background estimation methods are included in the systematic uncertainties. 
For weakly decaying particles, topological cuts are varied as well. Different topological variables are varied simultaneously to keep the raw yield of the particle of interest  similar. This helps to reduce the effect of statistical fluctuations in estimating systematic uncertainties. Finally, the Barlow's method~\cite{barlow} is used to determine the systematic uncertainties arising from analysis cut variations. If the resulting changes ($\Delta v_{n}$) in $v_{n}$   are smaller than the change in statistical errors ($\Delta \sigma_{stat}$) on $v_{n}$, such changes are not included in the uncertainties. Otherwise, the systematic error ($\sigma_{sys}$) on $v_{n}$  is calculated as $\sigma_{sys}$ =$\sqrt{ (\Delta v_{n})^{2} -  (\Delta \sigma_{stat})^{2} }$.
Finally, systematic uncertainties from different sources, which pass the Barlow check, are added in quadrature.  Final systematic uncertainties are calculated as a function of $p_{T}$ and centrality. They are found to be nearly $p_{T}$ independent but larger in central collisions compared to peripheral collisions.
Table~\ref{tab_sys_v2} and \ref{tab_sys_v3} show the average systematic uncertainties on $v_{2}$ and $v_{3}$ for $K_{S}^{0}$, $\phi$, $\Lambda$, $\Xi$ and $\Omega$ in different centrality bins.

\begin{table}[]
\begin{tabular}{|l|l|l|l|l|}
\hline
 Particle/Centrality& 0-10\%  & 10-40\% & 40-80\% &  0-80\% \\ \hline
$K_{S}^{0}$ & 2\% &  2\% &  2\% & 2\% \\ \hline
$\phi$ & 10\%  & 3\%  & 3\%  & 5\% \\ \hline
$\Lambda$ &   2\%& 2\% &  2\% & 2\%  \\ \hline
$\Xi$ &  4\% &   3\%&  3\%&  3\% \\ \hline
$\Omega$ &   22\%&  6\% &  15\% & 8\%  \\ \hline
\end{tabular}
\caption{Average systematic uncertainties on $v_{2}$ of  $K_{S}^{0}$, $\phi$, $\Lambda$, $\Xi$ and $\Omega$ in different centrality bins.}
\label{tab_sys_v2}
\end{table}

\begin{table}[]
\begin{tabular}{|l|l|l|l|l|}
\hline
 Particle/Centrality& 0-10\%  & 10-40\% & 40-80\% &  0-80\% \\ \hline
$K_{S}^{0}$ &  3\% &  3\% &  3\% &  3\% \\ \hline
$\phi$ &  15\%  &  10\%  & N.A. &  10\% \\ \hline
$\Lambda$ &   3\%&  3\% &  3\% &  3\%  \\ \hline
$\Xi$ &  12\% &  10\%&  N.A. &  8\% \\ \hline
$\Omega$ &   30\%&   30\%  & N.A. &  30\%  \\ \hline
\end{tabular}
\caption{Average systematic uncertainties on $v_{3}$  of $K_{S}^{0}$, $\phi$, $\Lambda$, $\Xi$ and $\Omega$ in different centrality bins.}
\label{tab_sys_v3}
\end{table}

\section{Results and discussion}
\subsection{$p_{T}$ dependence of $v_{2}$ and $v_{3}$}
\begin{figure*}
\centering
\includegraphics[scale=0.75]{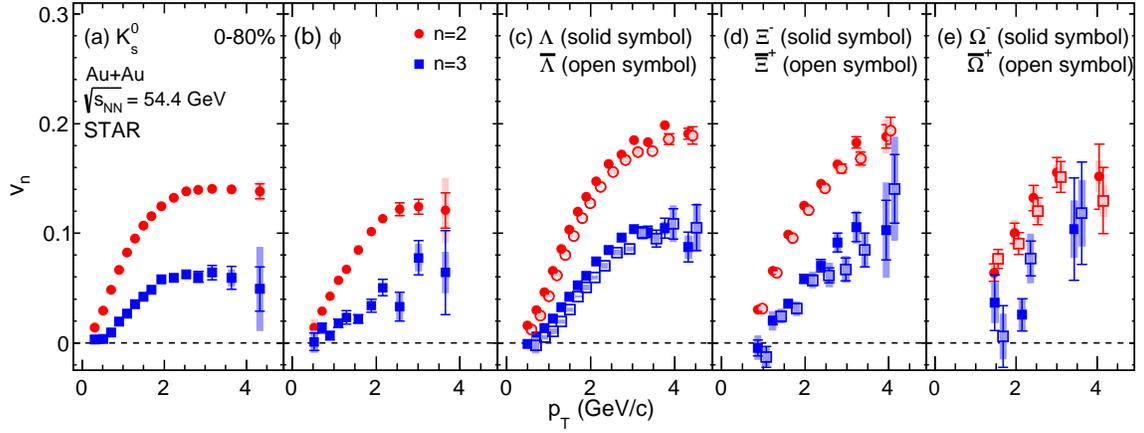}
\caption{$v_{2}$ and $v_{3}$ as a function of  $p_{T}$ at mid-rapidity($|y|<$1) for minimum bias events. The vertical lines represent the statistical error bars and the shaded bands represent the systematic uncertainties.
 Data points for  anti-particles are shifted by 0.1 GeV/c towards right for better visibility.
}
\label{fig:c}
\end{figure*}

The transverse momentum dependence of  $v_{2}$ and $v_{3}$ for  $K_{S}^{0}$, $\phi$, $\Lambda$, $\Xi^{-}$, $\Omega^{-}$ (and their anti-particles) is shown in Fig.~\ref{fig:c}. The measurements are done at mid-rapidity, $|y|$ $<$ 1.0, in  minimum bias Au+Au collisions at  $\sqrt{s_{\text{NN}}}$ = 54.4 GeV.  The non-zero magnitude of  $v_{3}$  is consistent with the picture of event-by-event fluctuations in the initial density profile of the colliding nuclei~\cite{alver}. 
Both $v_{2}$ and $v_{3}$ initially increase with $p_{T}$ and then tend to saturate. This may be due to the interplay of hydrodynamic flow as well as viscous effects~\cite{matt_hydro}.
The magnitude of $v_{3}$  is found to be less than that of $v_{2}$ for all particles in 0-80\% centrality. This is the first $v_{3}$ measurement of the multi-strange baryons $\Xi$ and $\Omega$  in relativistic heavy-ion collisions.
The $v_{n}$ of heavy multi-strange baryons like $\Omega$ are similar to that of the lighter mass, strange baryon $\Lambda$. The $v_{n}$ of $\phi$ mesons,  which consist of strange and anti-strange quark pairs, is similar to that of light, strange $K_{S}^{0}$. 
If $v_{n}$  is developed through hadronic interactions, $v_{n}$ should depend on the cross-sections of the interacting hadrons and therefore those (e.g. $\phi$, $\Omega$) with smaller cross-sections should develop less momentum anisotropy. Therefore the observed large $v_{n}$ of  $\phi$ and $\Omega$ are consistent with the scenario that the anisotropy is developed in the partonic medium in Au+Au collisions at  $\sqrt{s_{\text{NN}}}$ = 54.4 GeV.
We also observe a difference in $v_{n}$ between baryon and anti-baryon which is discussed separately in a later section. 
The high precision measurements of $v_{n}$ for $K_{S}^{0}$, $\phi$, $\Lambda$, $\Xi$,  and $\Omega$ presented in this paper can be used to constrain various models, for example, in extracting transport properties of the medium created  at  $\sqrt{s_{\text{NN}}}$ = 54.4 GeV.

\subsection{Centrality dependence of $v_{2}$ and $v_{3}$}
\begin{figure*}
\centering
\includegraphics[scale=0.75]{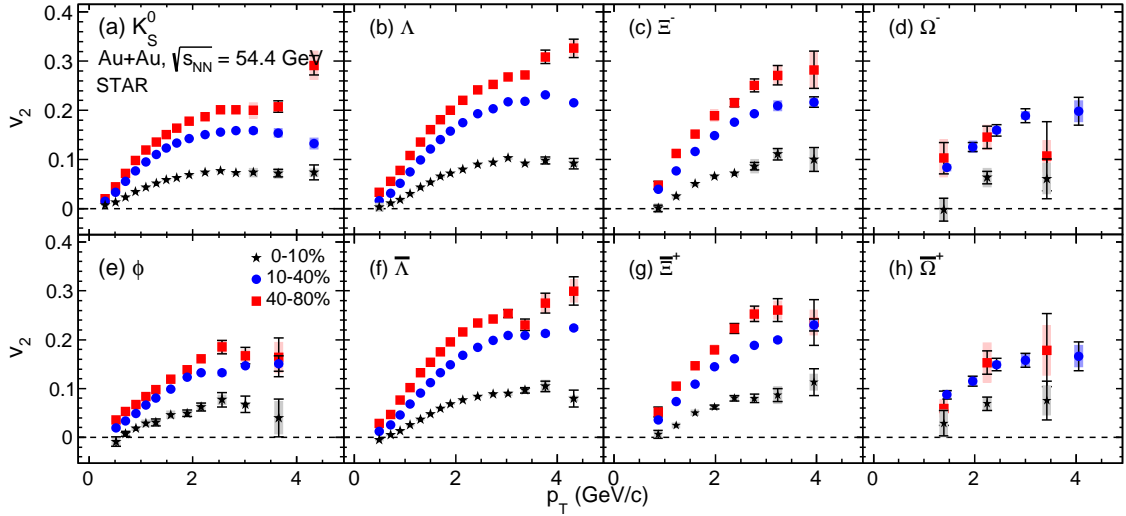}
\caption{$v_{2}$ as function of $p_{T}$ for 0-10\%, 10-40\% and 40-80\% centrality events. The vertical lines represent the statistical error bars and the shaded bands represent the systematic uncertainties.}
\label{fig:d}
\end{figure*}

\begin{figure*}
\centering
\includegraphics[scale=0.75]{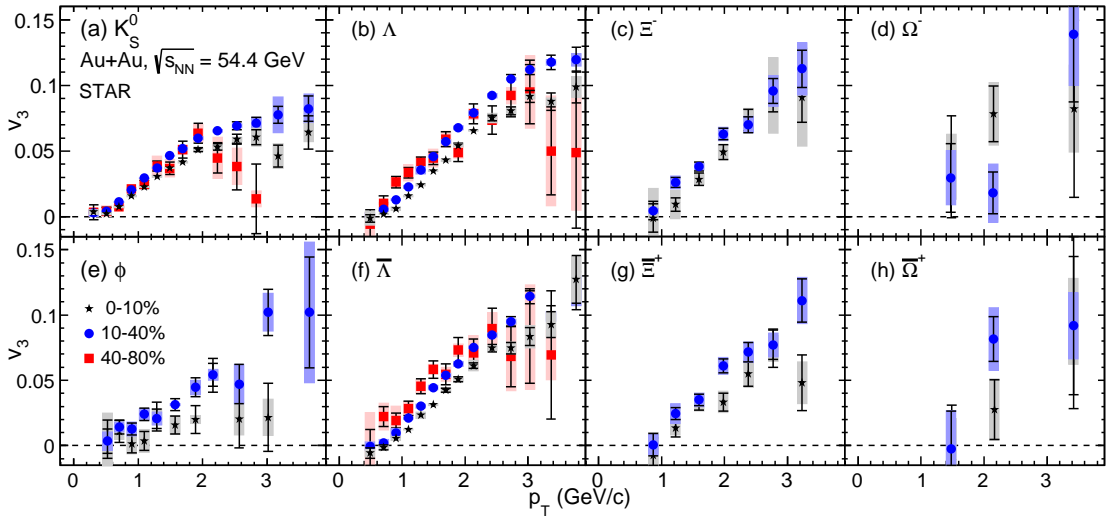}
\caption{$v_{3}$ as function of $p_{T}$ for 0-10\%, 10-40\% and 40-80\% centrality events. The vertical lines represent the statistical error bars and the shaded bands represent the systematic uncertainties. 40-80\% centrality data points are not shown for $\Xi$ and $\Omega$ due to less statistics.}
\label{fig:e}
\end{figure*}

The centrality dependence of $v_{2}$ and $v_{3}$ of  $K_{S}^{0}$, $\phi$, $\Lambda$, $\Xi^{-}$, $\Omega^{-}$ (and their anti-particles) are studied. Figures~\ref{fig:d} and ~\ref{fig:e} show $v_{2}$ and $v_{3}$, respectively, as a function of $p_{T}$ for three different centrality classes, 0-10\%, 10-40\% and 40-80\%. For $\phi$, $\Xi$ and $\Omega$ measurements are only possible for $v_{3}$ for the 0-10\% and 10-40\% centralities due to data sample size. We observe a strong centrality dependence of $v_{2}$ for all the particles, with the magnitude increasing from central to peripheral collisions.  This is expected if $v_{2}$ is driven by the shape of the initial overlap of the two colliding nuclei~\cite{star_bes_chg}. 
\par
We observe a weak centrality dependence for $v_{3}$ compared to $v_{2}$. This observation is consistent with the scenario in which $v_{3}$ mostly originates from event-by-event fluctuations of participant nucleon distributions~\cite{alver}, instead of the impact parameter dominated average participant anisotropy distributions. Our measurements demonstrate that such scenario also works well for 54.4 GeV Au+Au collisions.

\subsection{Energy dependence of $v_{2}$ and $v_{3}$}
\begin{figure*}[!htbp]
\centering
\includegraphics[scale=0.8]{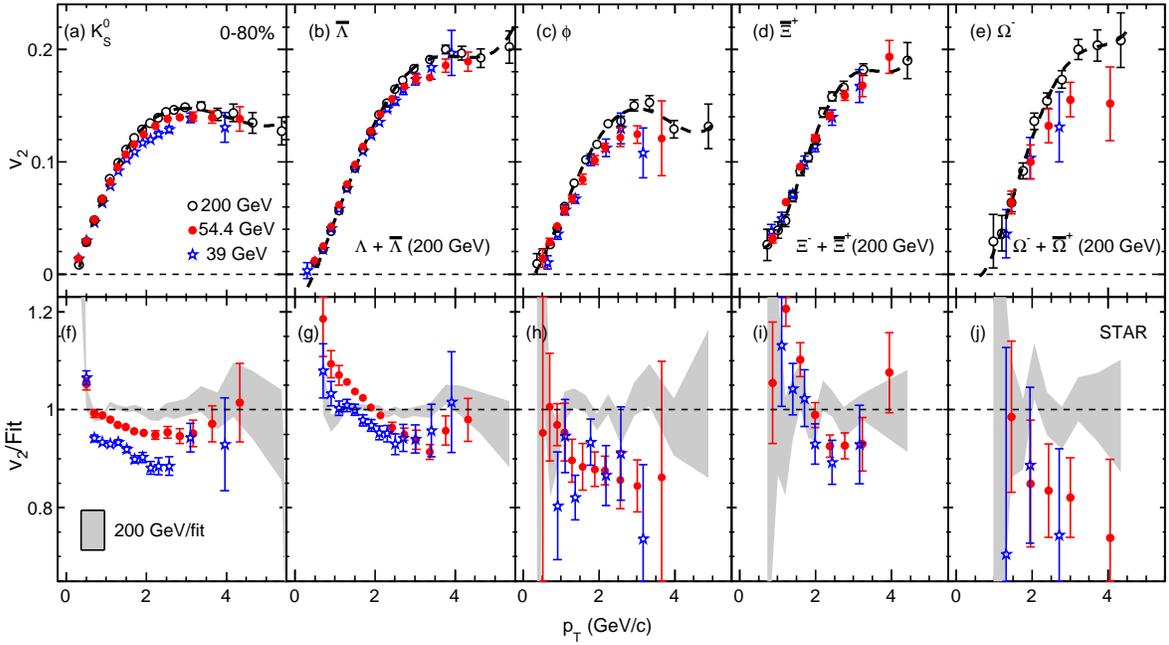}
\caption{$v_{2}$ of $K_{S}^{0}$, $\phi$, $\bar{\Lambda}$, $\bar{\Xi}^{+}$, and $\Omega^{-}$ as a function of $p_{T}$ in 0-80\% centrality events at $\sqrt{s_{\text{NN}}}$ = 39, 54.4, and 200 GeV. The dotted line represents the fit to the 200 GeV data points. The vertical lines represent the sum of statistical and systematic uncertainties in quadrature. The data points for 39 and 200 GeV are taken from refs.~\cite{star_200gev_prl,star_bes_prl,run4v2}}
\label{fig:energy_v2}
\end{figure*}

\begin{figure*}[!htbp]
\centering
\includegraphics[scale=0.7]{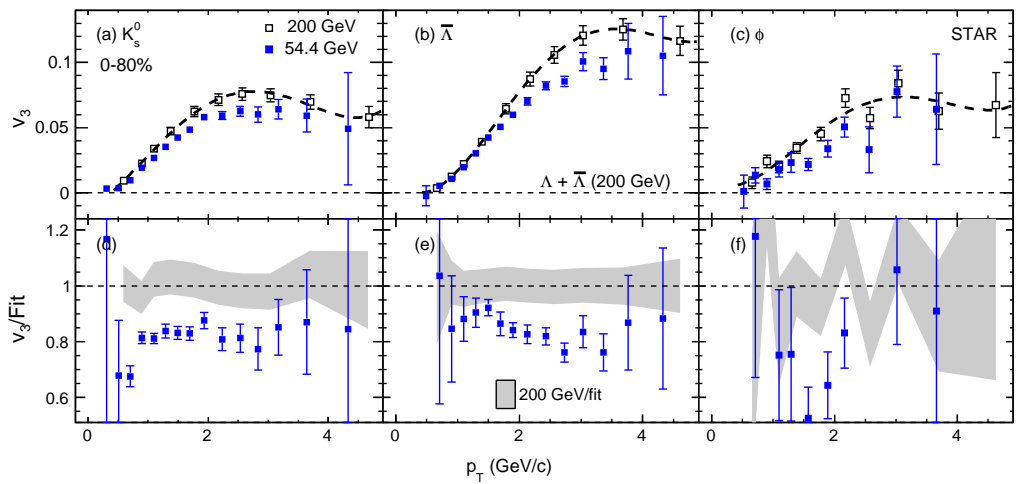}
\caption{$v_{3}$ of $K_{S}^{0}$, $\phi$ and $\bar{\Lambda}$ as a function of $p_{T}$ in 0-80\% centrality events at $\sqrt{s_{\text{NN}}}$ = 54.4,  and 200 GeV. The vertical lines represent the sum of statistical and systematic uncertainties in quadrature. The  data points for 200 GeV are taken from ref.~\cite{200gev_nissem}. }
\label{fig:energy_v3}
\end{figure*}

The high statistics data at 54.4 GeV from the STAR experiment offer an opportunity to study the collision energy dependence of $v_{2}$ and $v_{3}$ of strange hadrons. 
Figure~\ref{fig:energy_v2} upper panels show $v_{2}$ of $K_{S}^{0}$, $\phi$, $\bar{\Lambda}$, $\bar{\Xi}^{+}$, and $\Omega^{-}$ as a function of $p_{T}$ in 0-80\% centrality  at $\sqrt{s_{\text{NN}}}$ = 39, 54.4, and 200 GeV. Lower panels show the
ratios with polynomial fits to the 200 GeV data points. $K_{S}^{0}$ $v_{2}$ at 54.4 GeV is smaller than at 200 GeV, and higher than at 39 GeV. The maximum difference is at intermediate $p_{T}$.
For $\bar{\Lambda}$ and $\bar{\Xi}^{+}$, $v_{2}$ at 54.4 GeV (as well as at 39 GeV) is higher than at 200 GeV at very low $p_{T}$. This could be due to the effect of large radial flow at 200 GeV compared to 54.4 and 39 GeV. This effect is only visible in heavier hadrons like $\bar{\Lambda}$ and $\bar{\Xi}^{+}$.  For $\phi$ and  $\Omega^{-}$, statistical errors at low $p_{T}$ are too large to draw any conclusions.
Figure~\ref{fig:energy_v3} upper panels shows $v_{3}$ of $K_{S}^{0}$, $\phi$, and $\bar{\Lambda}$  as a function of $p_{T}$ in 0-80\% centrality  at $\sqrt{s_{\text{NN}}}$ = 54.4,  and 200 GeV. Lower panels show the ratios of fits to the 200 GeV data points. We observe that the difference in $v_{3}$ between 54.4 and 200 GeV is almost $p_{T}$ independent for all the particles studied. In Fig.~\ref{fig:energy_v3}, the $v_{3}$ shows greater variation as a function of beam energy than that of $v_{2}$. The measured ratio of $v_{3}$(54.4 GeV)/$v_{3}$(200 GeV) for $K_{S}^{0}$ is $\sim$0.8 while the same ratio for $v_{2}$ is approaching 0.9.
This suggests that the dynamics responsible for $v_{3}$, presumably fluctuations dominated, are more sensitive to beam energy than the $v_{2}$.

\subsection{$v_{n}$ of particles and anti-particles}
\begin{figure*}
\centering
\includegraphics[scale=0.7]{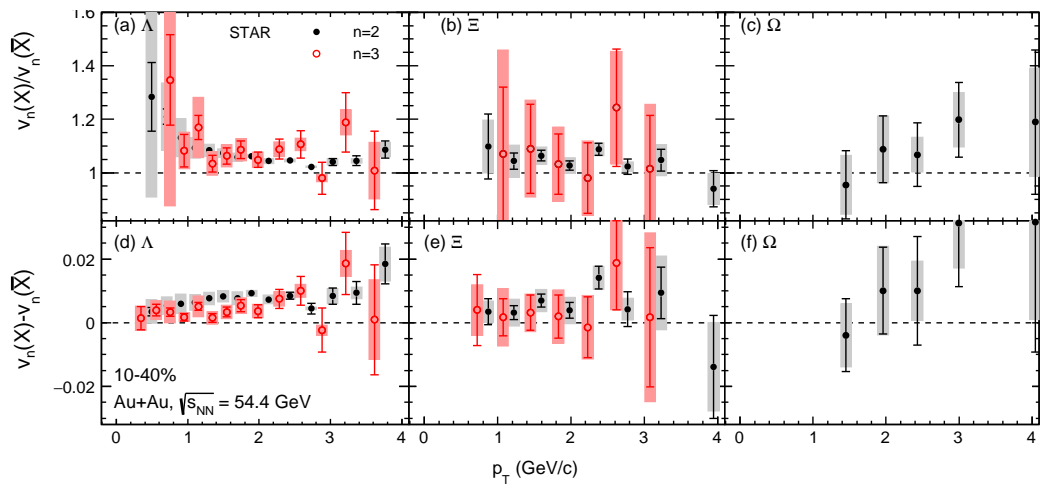}
\caption{Three upper panels, a, b, and c show the ratio of $v_{n}$ of particles to anti-particles for $\Lambda$, $\Xi$ and $\Omega$ respectively in 10-40\% centrality. The lower panels show the difference between  $v_{n}$ of particles to anti-particles. The vertical lines represent the statistical error bars and the shaded bands represent the systematic uncertainties. Data points for  $v_{3}$ are shifted by 0.15 GeV/c towards the left for better visibility.
For the $\Omega$, data points for $v_{3}$ were not shown due to fewer statistics.}
\label{fig:1}
\end{figure*}

\begin{figure}
\centering
\includegraphics[scale=0.4]{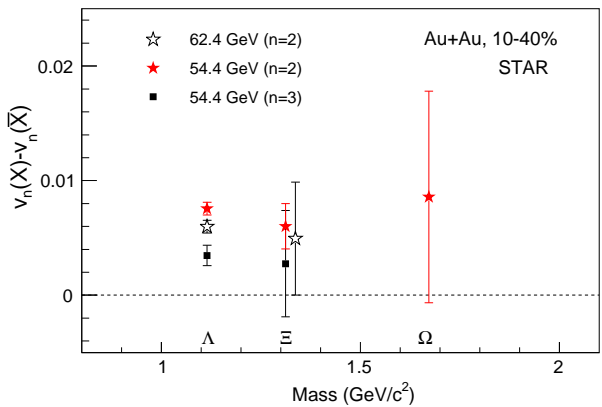}
\caption{The difference of $v_{n}$ of particles and anti-particles is plotted as a function of mass. The result is compared with 62.4 GeV. Uncertainties represent the sum of statistical and systematic in quadrature.}
\label{fig:2}
\end{figure}

In the upper panels Fig~\ref{fig:1}, we show the ratio of $v_{2}$ and $v_{3}$ of particles ($v_{n}(X)$) to the corresponding anti-particles ($v_{n}(\bar{X})$) for  $\Lambda$, $\Xi$, and $\Omega$ in 10-40\% centrality as a function of $p_{T}$. We also present the difference between $v_{2}$ and $v_{3}$ of particles and anti-particles in the lower panels of Fig .~\ref{fig:1}. We can not establish a clear $p_{T}$ dependence in the ratio or difference of multi-strange particle and anti-particle. The $\Lambda$ and $\bar{\Lambda}$ $v_{n}$ data seem to be consistent with a relatively smaller $v_{n}$ for $\bar{\Lambda}$ in the low $p_{T}$ region. We have calculated the $p_{T}$ integrated average difference in $v_{n}$ between baryon and anti-baryon by fitting the $v_{n}(X)-v_{n}(\bar{X})$ versus $p_{T}$ with a zeroth order polynomial function as done in Ref.~\cite{star_bes_prl}. Figure~\ref{fig:2} shows the average difference between $v_{n}$ of baryons and anti-baryons for $\Lambda$, $\Xi$, and $\Omega$ in 10-40\% centrality as a function of mass. The difference is independent of baryon species within the measured uncertainty for both $v_{2}$ and $v_{3}$. The magnitude of the observed difference between particle and anti-particle is similar to that in 62.4 GeV published by the STAR experiment~\cite{bes_pid_prc}.  However, uncertainties on the measured values are significantly reduced at 54.4 GeV. 
The observed difference between particles and anti-particles could arise due to the effect of transported quarks at low beam energies as predicted in ~\cite{jamie}. Alternatively, a calculation  based on  the Nambu-Jona-Lasinio (NJL) model~\cite{pnjl_1,pnjl_2} can also qualitatively explain the differences between particles and anti-particles by considering the effect of the vector mean-field potential, which is repulsive for quarks and attractive for antiquarks. We also measure the difference between  $\Omega^{-}$ and  $\bar{\Omega}^{+}$, however the observed difference is not statistically significant ($<$1$\sigma$ significance).

\subsection{$v_{3}$/$v_{2}^{3/2}$ ratio}
\begin{figure*}[!htbp]
\centering
\includegraphics[scale=0.7]{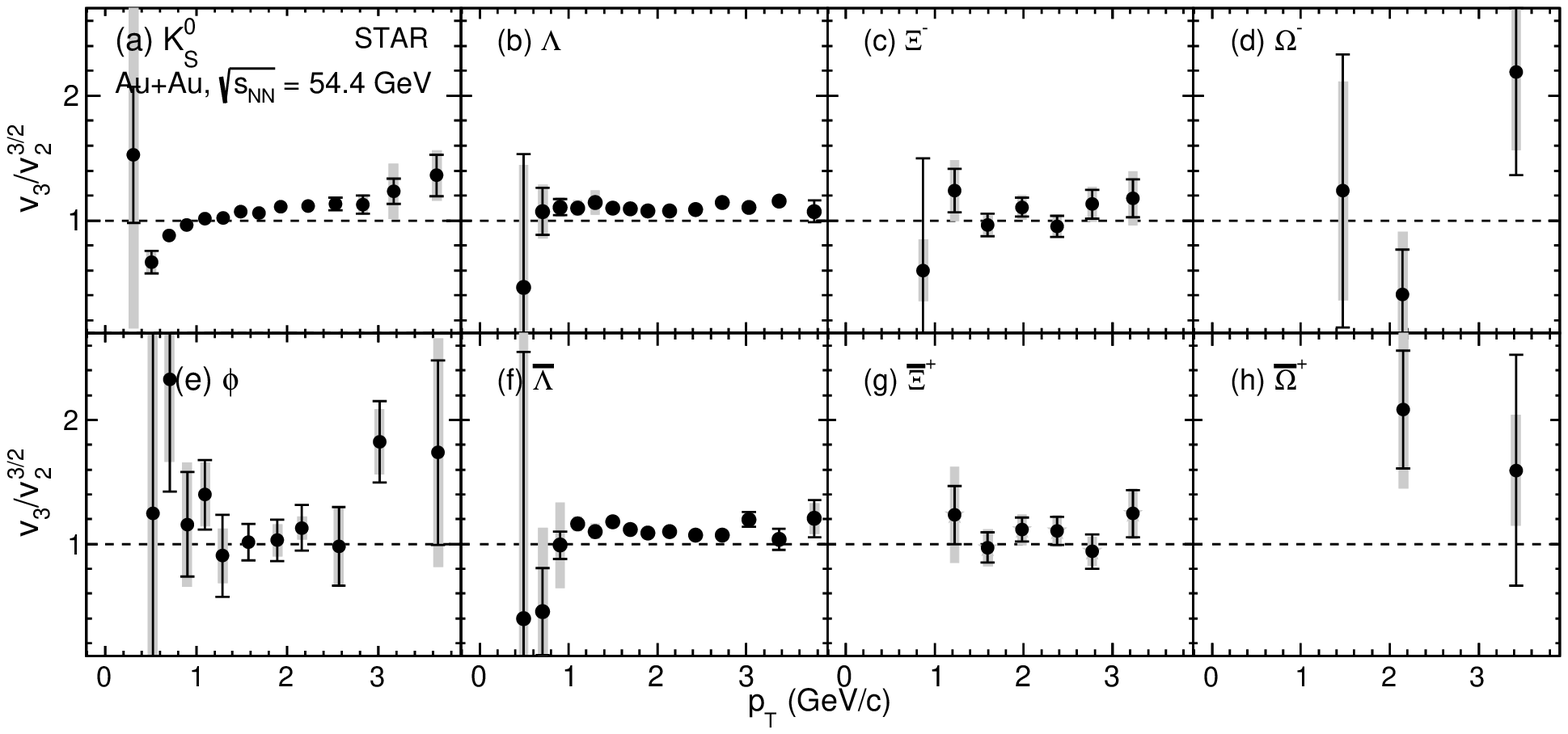}
\caption{$v_{3}$/$v_{2}^{3/2}$ is plotted as a function of $p_{T}$ for $K_{S}^{0}$, $\Lambda$, $\Xi^{-}$, $\Omega^{-}$, $\phi$, $\bar{\Lambda}$, $\bar{\Xi}^{+}$ and $\bar{\Omega}^{+}$ in 10-40\% central Au+Au collisions at $\sqrt{s_{\text{NN}}}$ = 54.4 GeV. }
\label{fig:h}
\end{figure*}

The ratios between different orders of flow harmonics are predicted to be sensitive probes of transport properties of the produced medium in heavy-ion collisions.  According to hydrodynamic model calculations, the ratio $v_{3}$/$v_{2}^{3/2}$ is independent of $p_{T}$ and its magnitude depends on the transport properties (e.g., viscosity) of the medium~\cite{hydro_v2_v3_1,hydro_v2_v3_2,hydro_v2_v3_3}.
We have calculated the ratio $v_{3}$/$v_{2}^{3/2}$ as a function of $p_{T}$ for $K_{S}^{0}$, $\Lambda$, $\Xi^{-}$, $\Omega^{-}$, $\phi$, $\bar{\Lambda}$, $\bar{\Xi}^{+}$ and $\bar{\Omega}^{+}$ for 10-40\% centrality, as shown in Fig.~\ref{fig:h}. Our measurement for $K_{S}^{0}$ clearly demonstrates a $p_{T}$ dependence of the ratio. The $p_{T}$ dependence of the ratios for $\Lambda$ is weak and ratios for other strange hadrons are limited by statistical errors.  Detailed comparisons with other RHIC measurements ~\cite{star_prl_v2_v4,phenix_prl_v2_v4} and with more hydrodynamic model calculations will shed more light on the dynamics.

\subsection{Number of constituent quark scaling of $v_{2}$ and $v_{3}$}
Elliptic flow measurements at top RHIC energy suggest that a strongly-interacting partonic matter is produced in Au+Au collisions~\cite{star_200gev_prl}. This conclusion is based in part on the observation that the elliptic flow for identified baryons and mesons when divided by the number of constituent quarks ($n_{q}$) is found to scale with the transverse kinetic energy of the particles.

\begin{figure*}[!htbp]
\centering
\includegraphics[scale=0.6]{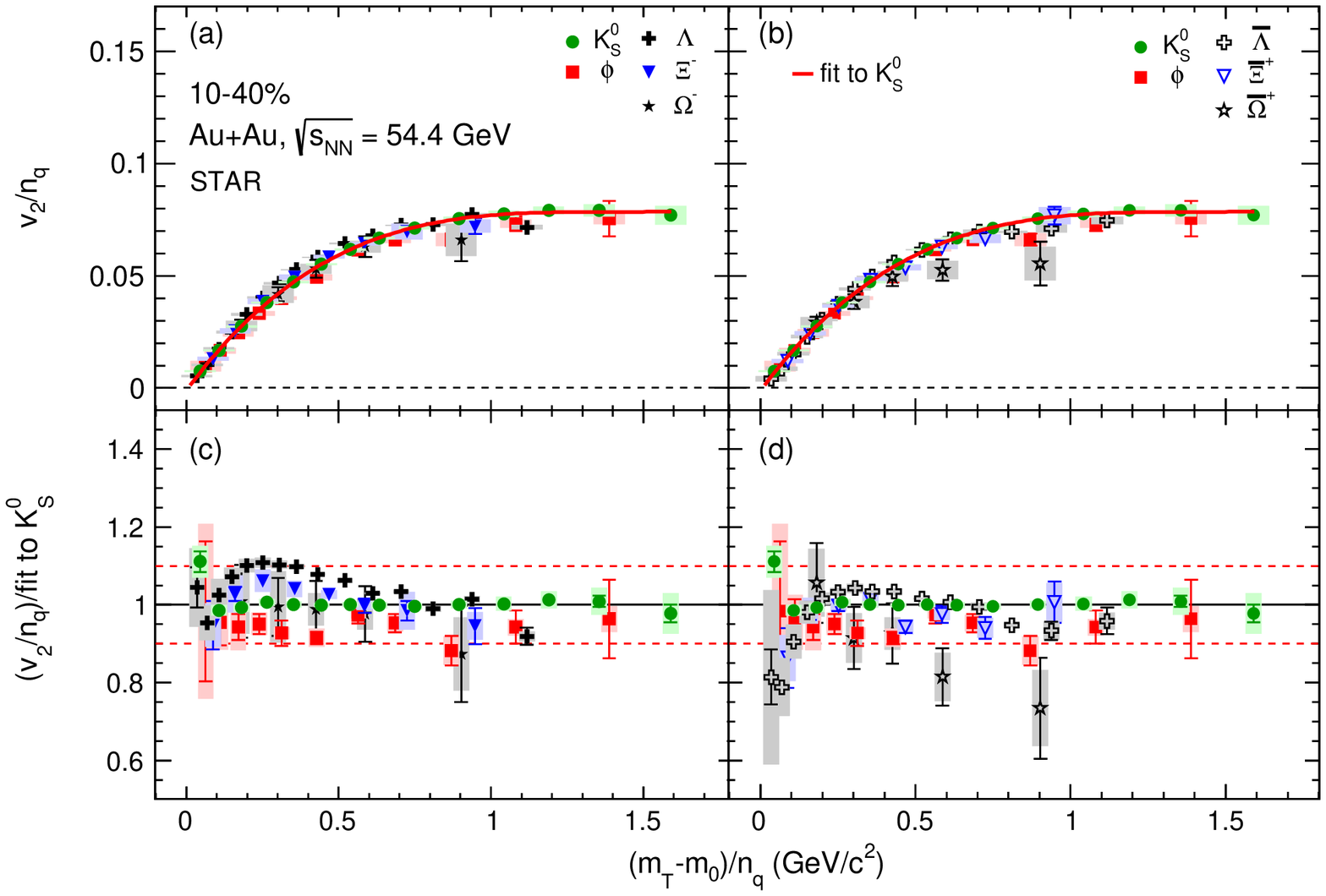}
\caption{Panel (a) shows the $n_{q}$-scaled $v_{2}$ as a function of $n_{q}$-scaled transverse kinetic energy for $K_{S}^{0}$, $\phi$, $\Lambda$, $\Xi^{-}$ and $\Omega^{-}$ in 10-40\% centrality class events. Panel (b) shows the same for $K_{S}^{0}$, $\phi$, $\bar{\Lambda}$, $\bar{\Xi}^{+}$ and $\bar{\Omega}^{+}$. The red line shows the polynomial fit to the $K_{S}^{0}$ data points. Panels (c) and (d) show the ratio of $n_{q}$-scaled $v_{2}$ of all the particles to the fit function. }
\label{fig:f}
\end{figure*}

Figure ~\ref{fig:f}(a) and (b) show the  $v_{2}$/$n_{q}$ as a function of $n_{q}$ scaled transverse kinetic energy in 10-40\% central Au+Au collisions at $\sqrt{s_{\text{NN}}}$ = 54.4 GeV. The transverse kinetic energy is $m_{T}$-$m_{0}$ where $m_{T}$ is the transverse mass given by $m_{T}$ = $\sqrt{m_{0}^{2} + p_{T}^{2}}$ and $m_{0}$ is the rest mass of the particle.  Due to the observed difference in particle and anti-particle $v_{n}$ we plot $v_{2}/n_{q}$ vs. ($m_{T}$-$m_{0}$)/$n_{q}$ for particle and anti-particle separately. The $n_{q}$-scaled $v_{2}$ for identified hadrons including multi-strange hadrons are found to scale with the scaled kinetic energy of the particles. To quantify the validity of scaling we have fitted the scaled $v_{2}$ of $K_{S}^{0}$ with a 4th order polynomial, and ratios to the fit for different particles have shown in lower panels of Fig.~\ref{fig:f}. It is found that the scaling holds within a maximum deviation of 10\% for all the particles.
The observed scaling in $v_{2}$ can be interpreted as due to the development of substantial collectivity in the partonic phase~\cite{ncq_theory}  and as evidence that coalescence is the dominant mechanism of particle production for the intermediate $p_{T}$ range.
\par
\begin{figure*}[!htbp]
\centering
\includegraphics[scale=0.6]{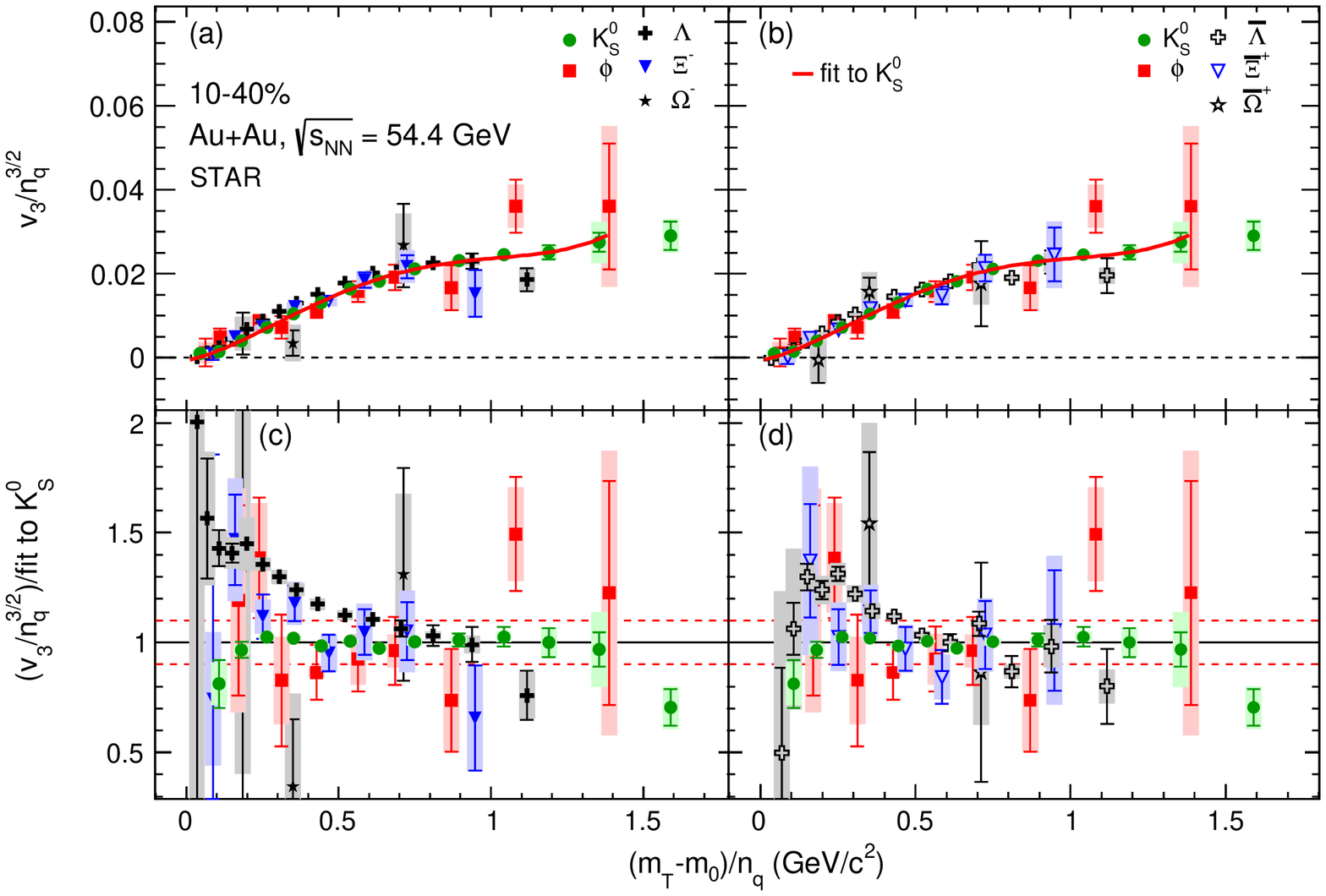}
\caption{Panel (a) shows $v_{3}/n_{q}^{3/2}$ as a function of $n_{q}$-scaled transverse kinetic energy for $K_{S}^{0}$, $\phi$, $\Lambda$, $\Xi^{-}$ and $\Omega^{-}$ in 10-40\% centrality class events. Panel (b) shows the same for $K_{S}^{0}$, $\phi$, $\bar{\Lambda}$, $\bar{\Xi}^{+}$ and $\bar{\Omega}^{+}$. The red line shows the polynomial fit to the $K_{S}^{0}$ data points. Panels (c) and (d) show the ratio of $v_{3}/n_{q}^{3/2}$ of all the particles to the fit function. }
\label{fig:g}
\end{figure*}

The scaling properties in $v_{3}$ have also been examined by plotting $v_{3}/(n_{q})^{3/2}$ as a function of ($m_{T}$-$m_{0}$)/$n_{q}$ as shown in panels (a) and (b) of Fig.~\ref{fig:g}. From the ratios shown in the lower panels, we note that the scaling of $v_{3}/(n_{q})^{3/2}$ is clearly violated for $\Lambda$ particles and the statistical errors for multi-strange particles are too large to draw a conclusion regarding scaling.

\subsection{$v_{2}(\phi)/v_{2}(\bar{p})$ ratio}
Among many mesons, the $\phi$($s\bar{s}$) has unique properties. It has a mass of 1.019 GeV/c$^{2}$ which is comparable to the mass of the lightest baryon, the proton (0.938 GeV/c$^{2}$). 
A hydrodynamical inspired study of transverse momentum distribution of $\phi$ meson seems to suggest that it freezes out early compared to other hadrons such as the proton~\cite{early_freezeout}.
Therefore, the kinematic properties of $\phi$ are expected to be less affected by the later stage hadronic interactions compared to the proton.

Hydrodynamical model calculations predict that $v_{2}$ of identified hadrons as a function of $p_{T}$ will follow mass ordering, where the $v_{2}$ of lighter hadrons is higher than that of heavier hadrons. A phenomenological calculation~\cite{hirano}, based on ideal hydrodynamics together with a hadron cascade (JAM), shows that because of late-stage hadronic rescattering effects  on the proton,  the mass ordering in $v_{2}$ will be violated between  $\phi$ and proton at very low $p_{T}$. This model calculation was done by assuming a small hadronic interaction cross-section for the $\phi$ meson and a larger hadronic interaction cross-section for protons, which is likely true for scatterings off the most abundant pions in the final state. However, several experimental and theoretical works on the
$\phi$-nucleon interaction that suggest that the magnitude of the cross section may not be negligible and more quantitative evaluations will be needed~\cite{phi_N_Coxy,phi_N_Coxy2,phi_N_ALICE,phi_N_KEK,phi_N_CLAS,phi_N_HADES,phi_N_theory,phi_N_theory2,phi_N_theory3}.

\begin{figure*}[!htbp]
\centering
\includegraphics[scale=0.7]{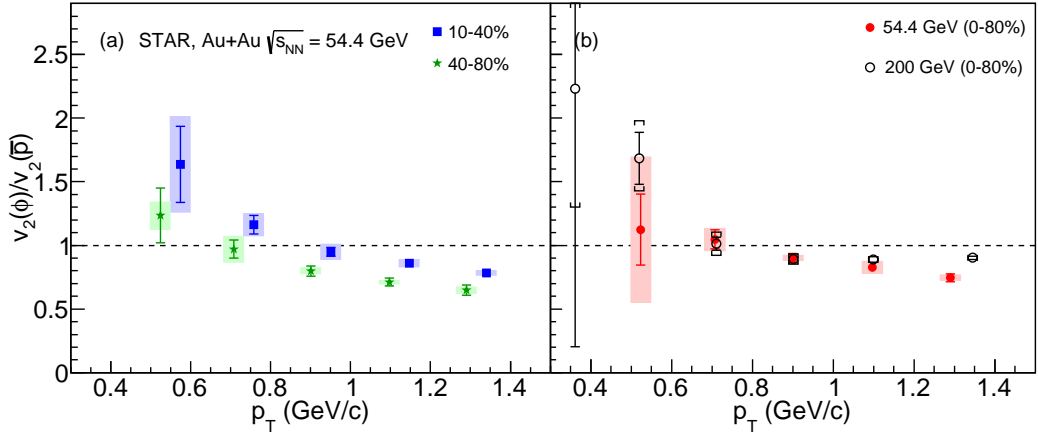}
\caption{Left panel shows the ratio of $v_{2}$ of $\phi$ to $v_{2}$ of $\bar{p}$ as a function of $p_{T}$ for 10-40\% and 40-80\% centralities at $\sqrt{s_{\text{NN}}}$ = 54.4 GeV. Data points for  10-40\% centrality are shifted by 0.05 GeV/c to the right for better visibility.
The right panel shows the comparison of the ratio at  $\sqrt{s_{\text{NN}}}$ = 54.4 GeV and 200 GeV in 0-80\% centrality. For 200 GeV~\cite{star_200gev_prl}, the measured ratio is  $v_{2}(\phi)/v_{2}(p+\bar{p})$  .The vertical lines represent the statistical error bars and the shaded bands represent the systematic uncertainties.
Data points at 200 GeV are taken from ref.~\cite{star_200gev_prl}}
\label{fig:i}
\end{figure*}

The breaking of mass ordering in $v_{2}$ between $\phi$ and proton was observed  in central Au+Au collisions at $\sqrt{s_{\text{NN}}}$ = 200 GeV and reported by the STAR experiment in Ref.~\cite{star_200gev_prl}. 
Figure~\ref{fig:i}(a) shows $v_{2}(\phi)/v_{2}(\bar{p})$ vs. $p_{T}$  for 10-40\% and 40-80\% centralities at $\sqrt{s_{\text{NN}}}$ = 54.4 GeV. The result for 0-10\% is not shown due to very large uncertainties. Anti-protons, which consist of all produced quarks ($\bar{u}\bar{u}\bar{d}$), are used instead of protons to avoid the effect of transported quarks.
At $p_{T}$ =0.5 GeV/c, the ratio is greater than one with 1$\sigma$ significance in 10-40\% centrality.  In addition, $v_{2}(\phi)/v_{2}(\bar{p})$ ratios in  10-40\% central collisions are found to be systematically higher than in peripheral 40-80\% events. This observed centrality dependence is consistent with the scenario of significant hadronic rescattering effect on $v_{2}$ of $\bar{p}$ while the effect for $\phi$ is considerably smaller~\cite{small_x_1,phiX2}. 
Comparison of the ratios for 0-80\% collision centrality from $\sqrt{s_{\text{NN}}}$ = 54.4 GeV and 200 GeV shows consistency with each other within uncertainties for $p_{T}$ $<$ 1.0 GeV/c.


\section{Summary}
In summary, we have reported the azimuthal anisotropic flow parameters, $v_{2}$ and $v_{3}$, of strange and multi-strange hadrons, $K_{S}^{0}$, $\phi$, $\Lambda$, $\Xi^{-}$, $\Omega^{-}$ (and their anti-particles) measured at mid-rapidity as a function of $p_{T}$ for various collision centralities in Au+Au collisions at $\sqrt{s_{\text{NN}}}$ =  54.4 GeV. The magnitude of $v_{3}$ of multi-strange baryons $\Xi$ and $\Omega$ is found to be similar to that of the lighter strange baryon $\Lambda$. The non-zero magnitude of $v_{3}$ indicates the presence of event-by-event fluctuations in the initial energy density profile of colliding nuclei and large values of $v_{2}$ and $v_{3}$ of  multi-strange hadrons indicate that the observed collectivity is mainly developed through partonic rather than hadronic interactions.
\par
The centrality dependence of $v_{3}$ is weak relative to that of $v_{2}$ which is consistent with the scenario that $v_{3}$ does not arise from impact parameter driven average spatial configurations, rather it originates dominantly from event-by-event fluctuation present in the system. 
The measured $v_{2}$ and $v_{3}$ values at $\sqrt{s_{\text{NN}}}$ = 54.4 GeV are also compared with available published results in Au+Au collisions at $\sqrt{s_{\text{NN}}}$ = 39 and 200 GeV to examine the energy dependence. We observed that the change in $v_{3}$ with $\sqrt{s_{\text{NN}}}$ is more than that in $v_{2}$. This suggests that $v_{3}$ dynamics have stronger energy dependence compared to $v_{2}$.
A difference in $v_{n}(p_{T} )$ between baryons and corresponding antibaryons was observed. The observed difference is found to be baryon-type independent within uncertainties.
\par
We have studied the $n_{q}$ scaling for both $v_{2}$ and $v_{3}$ and found that the scaling holds for $v_{2}$ of all the particles while the scaling for $v_{3}$ seems to be violated. One interpretation of the observed $n_{q}$ scaling in $v_{2}$ is that parton recombination is the dominant mechanism for hadronization at mid-rapidity and the development of collectivity occurs during the partonic stage of the system evolution. The ratio $v_{3}/v_{2}^{3/2}$, which is sensitive to the medium properties according to hydrodynamic calculations, shows weak $p_{T}$ dependence for $p_{T}>1$ GeV/c, similar to the behaviour of this ratio was found in the previous study with U+U collisions at 193 GeV. The $v_{2}(\phi)/v_{2}(\bar{p})$ ratio was presented as a function of $p_{T}$ for two different centrality classes 10-40\% and 40-80\%.  
The $v_{2}(\phi)/v_{2}(\bar{p})$ ratio shows a decreasing trend as a function of $p_{T}$ for both collision centralities.
The $v_{2}(\phi)/v_{2}(\bar{p})$ ratio is also found to be systematically higher for central collisions 10-40\% than non-central collisions 40-80\%.  This could be due the effect of more hadronic rescattering on $v_{2}$ of $\bar{p}$ compared to $\phi$ and hence our measurements are consistent with the picture of smaller hadronic rescattering and earlier freeze out of the $\phi$ mesons.


\begin{acknowledgements}
We thank the RHIC Operations Group and RCF at BNL, the NERSC Center at LBNL, and the Open Science Grid consortium for providing resources and support.  This work was supported in part by the Office of Nuclear Physics within the U.S. DOE Office of Science, the U.S. National Science Foundation, National Natural Science Foundation of China, Chinese Academy of Science, the Ministry of Science and Technology of China and the Chinese Ministry of Education, the Higher Education Sprout Project by Ministry of Education at NCKU, the National Research Foundation of Korea, Czech Science Foundation and Ministry of Education, Youth and Sports of the Czech Republic, Hungarian National Research, Development and Innovation Office, New National Excellency Programme of the Hungarian Ministry of Human Capacities, Department of Atomic Energy and Department of Science and Technology of the Government of India, the National Science Centre and WUT ID-UB of Poland, the Ministry of Science, Education and Sports of the Republic of Croatia, German Bundesministerium f\"ur Bildung, Wissenschaft, Forschung and Technologie (BMBF), Helmholtz Association, Ministry of Education, Culture, Sports, Science, and Technology (MEXT) and Japan Society for the Promotion of Science (JSPS).

\end{acknowledgements}

\end{document}